\documentclass[%
 reprint,
 amsmath,amssymb,
 pra,
]{revtex4-2}

\usepackage{graphicx}
\usepackage{dcolumn}
\usepackage{bm}
\usepackage{cases}
\usepackage{color}

\usepackage{blindtext}
\usepackage{float}
\usepackage{lipsum}

\begin{document}


\title{Thermal-induced Local Imbalance in Repulsive Binary Bose Mixtures}
\author{G. Pascual$^1$, G. Spada$^2$, S. Pilati$^{3,4}$, S. Giorgini$^2$ and J. Boronat$^1$}

\affiliation{
$^1$ Departament de F\'{i}sica, Universitat Polit\`ecnica de Catalunya, Campus Nord B4-B5, E-08034, Barcelona, Spain \\
$^2$ Dipartimento di Fisica, Universit\`a di Trento and CNR-INO BEC Center, 38123 Povo, Trento, Italy\\
$^3$ School of Science and Technology, Physics Division, Universit\`a di Camerino, 62032 Camerino, Italy\\
$^4$ INFN-Sezione di Perugia, 06123 Perugia, Italy
}

\date{\today}

\begin{abstract}
We study repulsive two-component Bose mixtures with equal populations and confined in a finite-size box through path-integral Monte Carlo simulations. For different values of the $s$-wave scattering length of the interspecies potential, we calculate the local population imbalance in a region of fixed volume inside the box at different temperatures. We find two different behaviors: for phase-separated states at $T=0$, thermal effects induce a diffusion process which reduces the local imbalance whereas, for miscible states at $T=0$, a maximum in the local population imbalance appears at a certain temperature, below the critical one. We show that this intriguing behavior is strongly related to the bunching effect associated with the Bose-Einstein statistics of the particles in the mixture and to an unexpected behavior of the cross pair distribution function not reported before.
\end{abstract}

\maketitle

\noindent \textit{Introduction.}
The experimental realization of quantum Bose-Bose mixtures with dilute gases~\cite{Myatt1997,Stamper1998,Stenger1998,Modugno2002,Thalhammer2008} has provided renewed interest on their theoretical study. Until this achievement, the only stable quantum mixture was the Fermi-Bose mixture composed by liquid $^4$He and $^3$He~\cite{Ebner70}. The Fermi nature of $^3$He atoms was here crucial to understand the solubility observed in experiments~\cite{Fabro82} and, in fact, it was proved theoretically that a fictitious Bose-Bose $^3$He-$^4$He mixture would be always unstable against phase separation~\cite{Tapash}. With ultracold gases, the high tunability of interactions and the possibility of mixing species with different masses allow the exploration of full phase diagrams, in both the miscible and immiscible regimes. Recent work on these quantum mixtures has led to compelling findings such as the discovery of quantum droplets (for an attractive interspecies interaction)~\cite{Petrov2015,Cabrera2018,Semeghini2018} or the occurrence of demixing phase transitions (for a repulsive interspecies interaction)~\cite{Esry1997,Pu1998,Ao1998,Timmermans1998,Trippenbach2000,Pethick2001,McCarron2011,Wacker2015,Wang2016,Lee2018}. 

In the case of repulsive interactions between all the particles of the mixture, a mean-field (MF) analysis at zero temperature shows that above the threshold $g_{12} = \sqrt{g_{11}g_{12}}$, where $g_{11}$, $g_{22}$, and $g_{12}$ are respectively the intraspecies and interspecies coupling constants, the two components of the mixture are phase separated  and, below that threshold, i.e., $g_{12}<\sqrt{g_{11}g_{22}}$, they are mixed~\cite{Ao1998}. A recent path-integral Monte-Carlo (PIMC) study shows that the same condition on the interaction coupling constants holds also at finite temperature, distinguishing fully miscible from partially phase separated states~\cite{Spada2022}. In particular, this result rules out the possibility of a paramagnetic to ferromagnetic transition occurring with increasing temperature, which was predicted using beyond mean-field perturbative theories~\cite{Ota2019,Ota2020}. Nonetheless, the behavior of Bose mixtures at finite temperature is a very interesting topic where effects from interactions and statistics combine producing an intriguing multicomponent superfluid phase.

The purpose of the present Letter is to analyze the thermal behavior of a Bose-Bose mixture in a confined environment. To this end, we use the PIMC method which is able to generate exact results for the thermodynamic properties of the system within controllable statistical errors. We use a box geometry like the one used in some recent experiments~\cite{Navon_naturephysics}. Our results for the local population imbalance show that its thermal behavior, below the Bose-Einstein transition temperature ($T_{\text{BEC}}$), is manifestly different for the states that at $T = 0$ are mixed or phase separated. When the ground state of the system is phase separated we observe a continuous tendency to mix and a corresponding reduction of the population imbalance, following a classical behavior with temperature. On the contrary, the local imbalance of zero-temperature mixed system shows in all cases a non monotonous behavior, with an intriguing thermally-induced increase up to a characteristic temperature (below $T_{\text{BEC}}$) from which the classical mixing mechanism prevails again.

\noindent \textit{Model.}
We describe a system of two different bosons with total number of particles $N=N_1+N_2$ in a cubic box of fixed volume $V$ using the following microscopic Hamiltonian,
\begin{eqnarray}\label{eq:Hamiltonian}
    H = &-&\frac{\hbar^2}{2m_1}\sum_{i=1}^{N_1}\nabla^2_i - 
\frac{\hbar^2}{2m_2}\sum_{i'=1}^{N_2}\nabla^2_{i'} + \sum_{i<j}^{N_1} V(r_{ij}) 
\nonumber\\
    &+& \sum_{i'<j'}^{N_2} V(r_{i' j'}) + \sum_{i,i'}^{N_1,N_2} 
V_{12}(r_{ii'}) \ ,
\end{eqnarray}
$m_1$, $m_2$ being the masses of the two species and particle indexes $i$ and $i'$ indicate the coordinates of particles of species 1 and 2, respectively.  We assume the same intraspecies potential for both components, $V(r)$, and study the influence on the properties of the mixture when the interspecies interaction $V_{12}(r)$ is changed. All the potentials are fully repulsive and we use a continuous model of the form $V(r)=(\alpha/r)^{12}$ and $V_{12}=(\beta/r)^{12}$. The corresponding $s$-wave scattering lengths $a$ and $a_{12}$ can be determined analytically from the parameters $\alpha$ and $\beta$ respectively~\cite{Pilati2006,Landau1977}. Throughout this work we use $a$ as the unit of length. Furthermore, our calculations are restricted to values of the gas parameter $na^3$ within the universal regime, in which the specific shape of the potential does not play any role~\cite{Giorgini1999}. To reduce the number of variables of our study we consider all particles with the same mass $m=m_1=m_2$ and the same number of particles for the two species $N_1=N_2=N/2$. The hard-wall conditions are imposed by rejecting any possible move of the particles outside the box. In the universal regime, the behavior of the mixture is only a function of the density, the strengths $g=\frac{4\pi\hbar^2a}{m}$ and $g_{12}=\frac{4\pi\hbar^2a_{12}}{m}$, and of the temperature $T$. In particular, temperature is given in units of 
\begin{equation}
T_{\text{BEC}} = 
\frac{2\pi\hbar^2}{m}\left(\frac{n}{2\zeta(3/2)}\right)^{2/3} \ , 
\label{tbec}
\end{equation}
where $n=N/V$ is the overall density in the box and $\zeta(x)$ is the Riemann zeta function.The above temperature corresponds to the Bose-Einstein condensation (BEC) in a non-interacting Bose gas at the uniform density $n/2$.

\noindent \textit{Method.}
We have used the PIMC method to calculate in an exact way, within controllable statistical noise, the microscopic  properties of the mixture at a given temperature. This method consists of dividing the thermal density matrix at some fixed $T$ into multiple density matrices (called \textit{beads}) at higher temperature, that can be well approximated~\cite{Ceperley1995}. In our work, we build the density matrix using the fourth-order Chin action~\cite{Chin2002}, whose accuracy has been validated in applications to other quantum systems~\cite{Sakkos2009}. The simulations need to include the Bose symmetric statistics of the particles. To this end, we sample the permutation space using the worm algorithm~\cite{Boninsegni2006}. To guarantee the distinguishability between particles of different species we have introduced the sampling of a second worm, in the same way as it has been done in  previous works~\cite{Pascual2021,Spada2022}.

\begin{figure}[t]
	\centering
	\includegraphics[width=5.2cm,height=4.2cm]{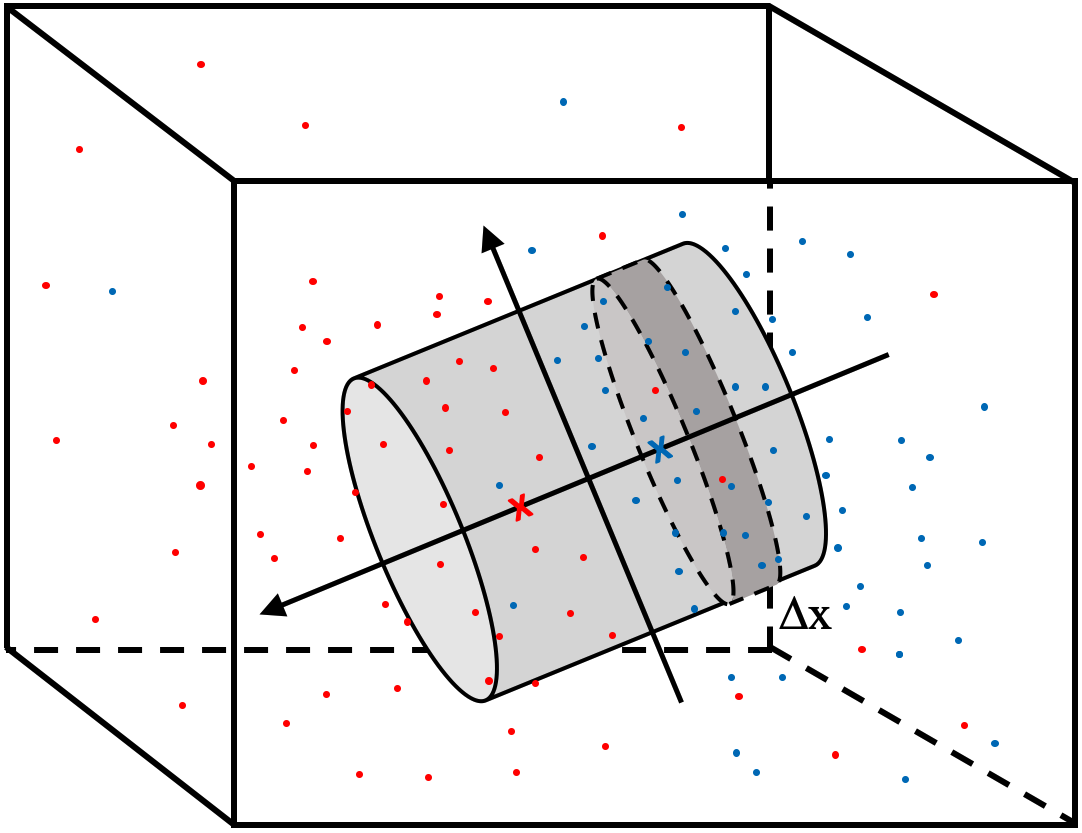}
	\caption{Schematic view of the method used to estimate the density profiles. The cylinder has the same diameter and length,  $L/\sqrt{2}$, where $L$ is the size of the cubic box.}
	\label{fig:cylinder}
\end{figure}

The sampling in the PIMC method is conducted in the coordinate space, therefore properties that depend on position operators can be easily calculated. However, as simulations are performed in a finite-size box, some technical issues must be taken into account. In particular, the problem of studying local population imbalance in a cubic box is that the system is degenerate, i.e., there is no privileged direction along which particles exhibit population imbalance or, possibly, phase separation. Therefore, since the Monte Carlo simulation of the finite system samples all possible configurations and, as a result, all possible degeneracies, the average of the density profiles of the mixture masks any possible imbalance between particles of different species. To avoid this effect and to sum constructively any possible configuration, we calculate the center of mass for each component and we define a one-dimensional density profile by integrating over particles along the axis joining the two centers of mass. In order to sample always the same slice of volume along the preferential axis, we only consider particles located inside a cylinder whose axis coincides with the axis connecting the two centers of mass. The size of this cylinder is fixed such that it does not exceed the limits of the box (diameter and length are fixed to $L/\sqrt{2}$, where $L$ is the size of the cubic box, see Fig.~\ref{fig:cylinder}). Different shapes and different sizes of the cylinder have been tested obtaining, for all of them, compatible results.

The local population imbalance is calculated as follows,
\begin{eqnarray}\label{deltaN}
    \frac{\delta\tilde{N}}{\tilde{N}} &=& \frac{1}{2}\int_{-L/(2\sqrt{2})}^{0}\frac{n_1(x)-n_2(x)}{n_1(x)+n_2(x)}dx \;+ \nonumber \\ &+& \frac{1}{2}\int_{0}^{L/(2\sqrt{2})}\frac{n_2(x)-n_1(x)}{n_1(x)+n_2(x)}dx,
\end{eqnarray}
where $\tilde{N}$ is the total number of particles inside the cylinder and the sign of the axis is chosen such that $n_2(x) > n_1(x)$ in the region $0<x<L/(2\sqrt{2})$.

\begin{figure}[H]
	\centering
	\includegraphics[width=8.6cm,height=4.5cm]{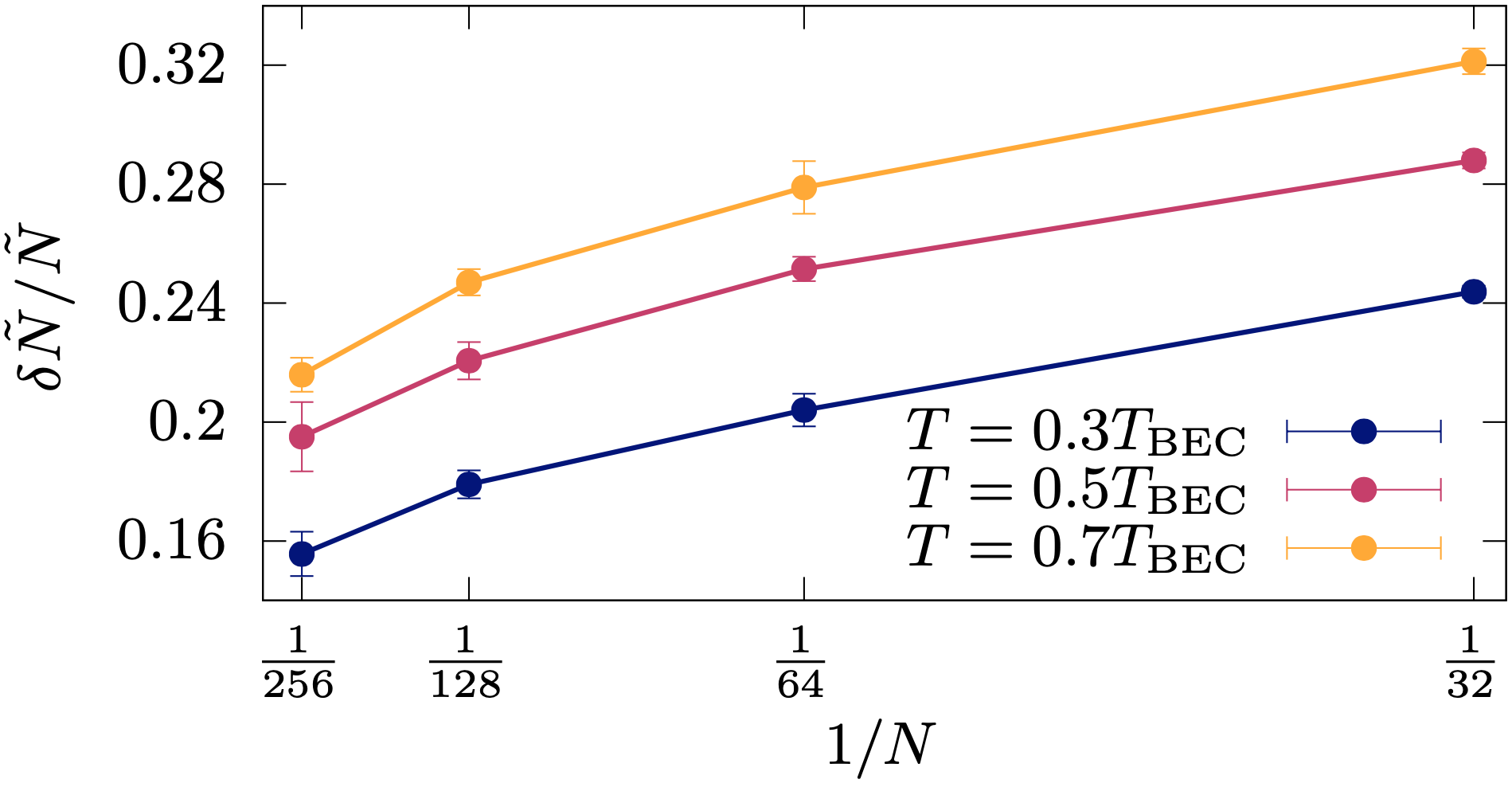}
	\caption{Local population imbalance of the mixture with respect to the total number of particles at $g_{12}/g = 0.93$ and $na^3 = 10^{-4}$.}
	\label{fig:ThermodynamicLimit}
\end{figure}

Apart from ensuring the convergence of the results as a function of the number of \textit{beads} used in the simulation, we analyze the imbalance $\delta\tilde{N}/\tilde{N}$ inside the cylinder with respect to the total number of particles in the box. In Fig. \ref{fig:ThermodynamicLimit}, we  show the dependence of $\delta\tilde{N}/\tilde{N}$ with the total number $N$ of particles in the box (at fixed overall density $N/V$). The results for the local imbalance decrease with increasing volume of the cylinder and are compatible with the expectation $\delta\tilde{N}/\tilde{N}=0$ holding in the thermodynamic limit for a paramagnetic mixture.

\noindent \textit{Results.}
We fix the gas parameter to $na^3 = 10^{-4}$ for all the simulations. With this choice, the system is dilute enough to be in the universal regime~\cite{Giorgini1999} and, at the same time, PIMC results converge faster than with smaller densities~\cite{Pascual2021}. 

\begin{figure}[b]
	\centering
	\includegraphics[width=8.6cm,height=12cm]{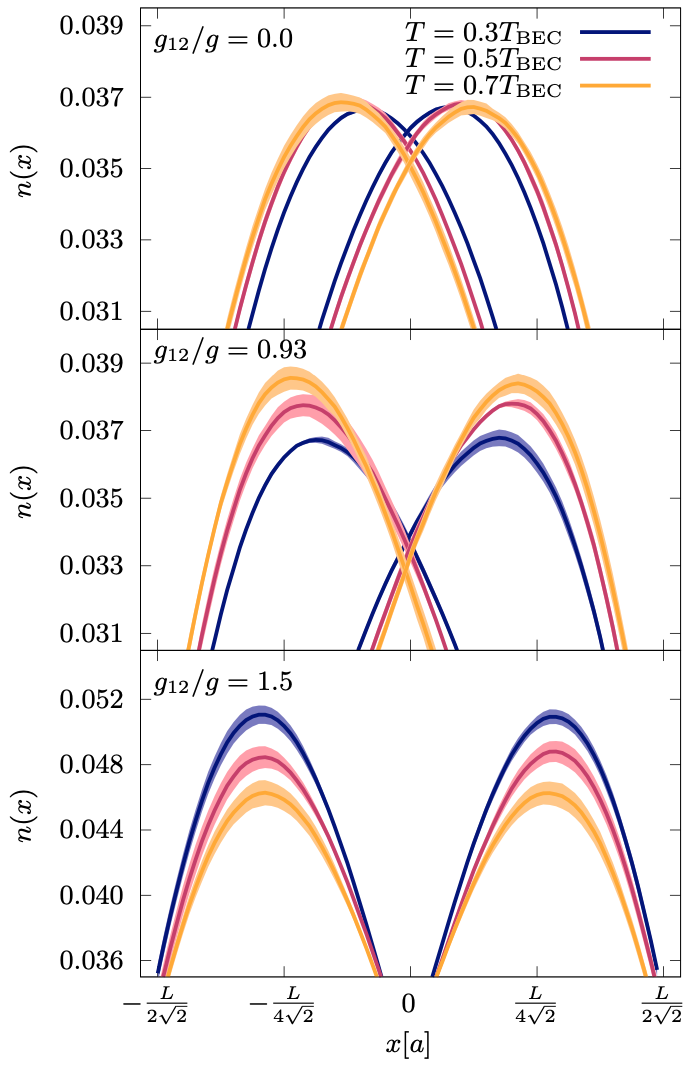}
	\caption{Density profiles at different temperatures for different values of $g_{12}/g$. The shaded area visualizes the statistical error obtained by averaging over different configurations.}
	\label{fig:DensityProfiles}
\end{figure}

In Fig.~\ref{fig:DensityProfiles}, we show the density profiles along the axis of the cylinder of the two species (each curve of the same color represents the profile of component 1 and 2) at three different temperatures and for different values of $g_{12}/g$. For values below the phase-separation threshold $g_{12}/g = 1$ (top and middle plot), there is a constant trend: the peak of the density profiles becomes narrower and more separated as the temperature is increased up to the maximum value $T = 0.7T_{\text{BEC}}$. Moreover, for larger $g_{12}$ values approaching the threshold this effect is enhanced. On the other hand, above  $g_{12}/g = 1$  (bottom panel) that is, when the system is phase separated at $T=0$, we see a clear suppression of the peak with increasing $T$ consistent with a thermal behavior.   

\begin{figure}[t]
	\centering
	\includegraphics[width=8.6cm,height=7.8cm]{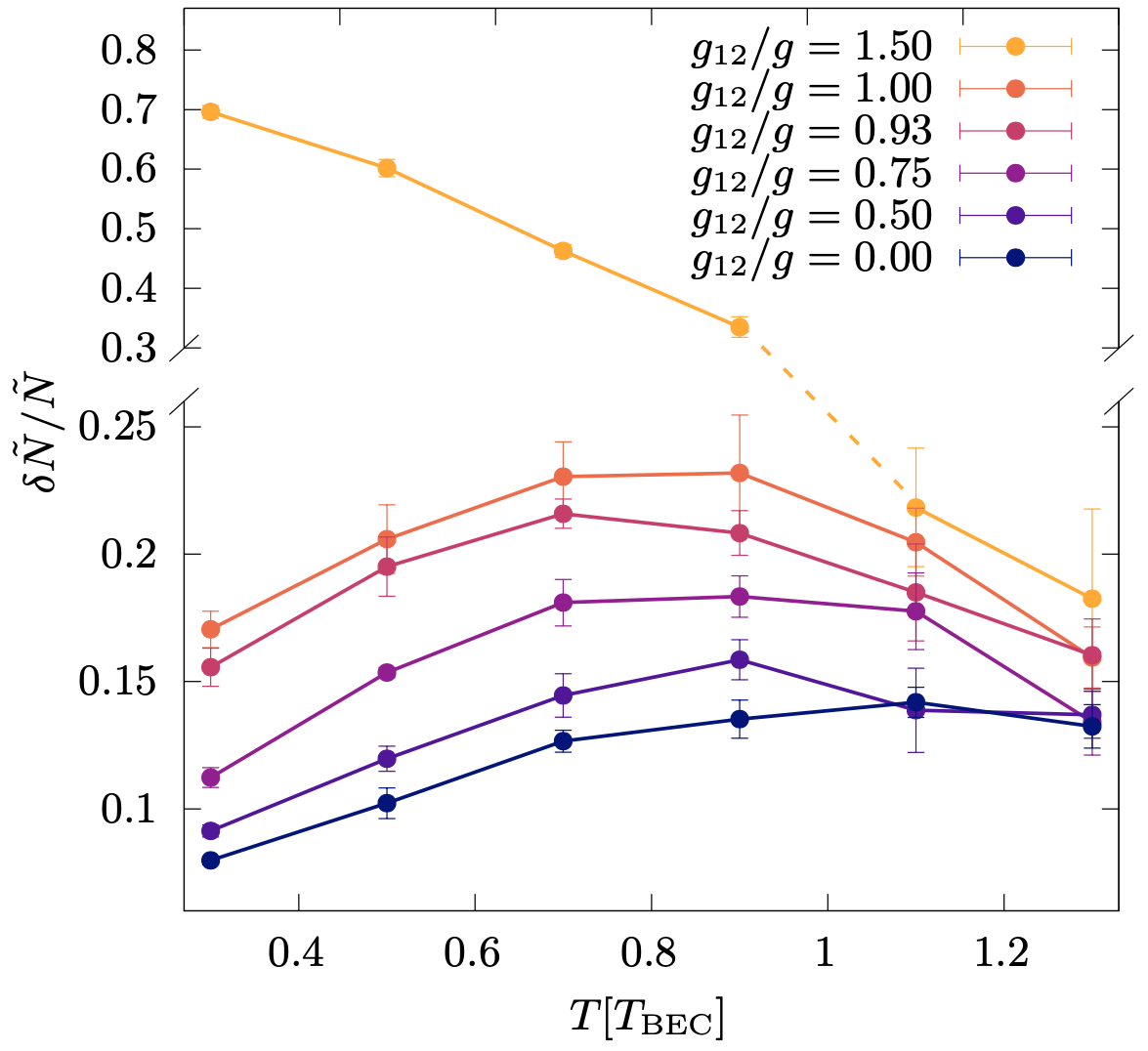}
	\caption{Local population imbalance as a function of temperature for different values of $g_{12}/g$ with $N=256$. Note that the vertical axis is broken in order to show the different scale when $g_{12}/g > 1$.}
	\label{fig:Imbalance}
\end{figure}

\begin{figure}[b]
	\centering
	\includegraphics[width=8.6cm,height=5cm]{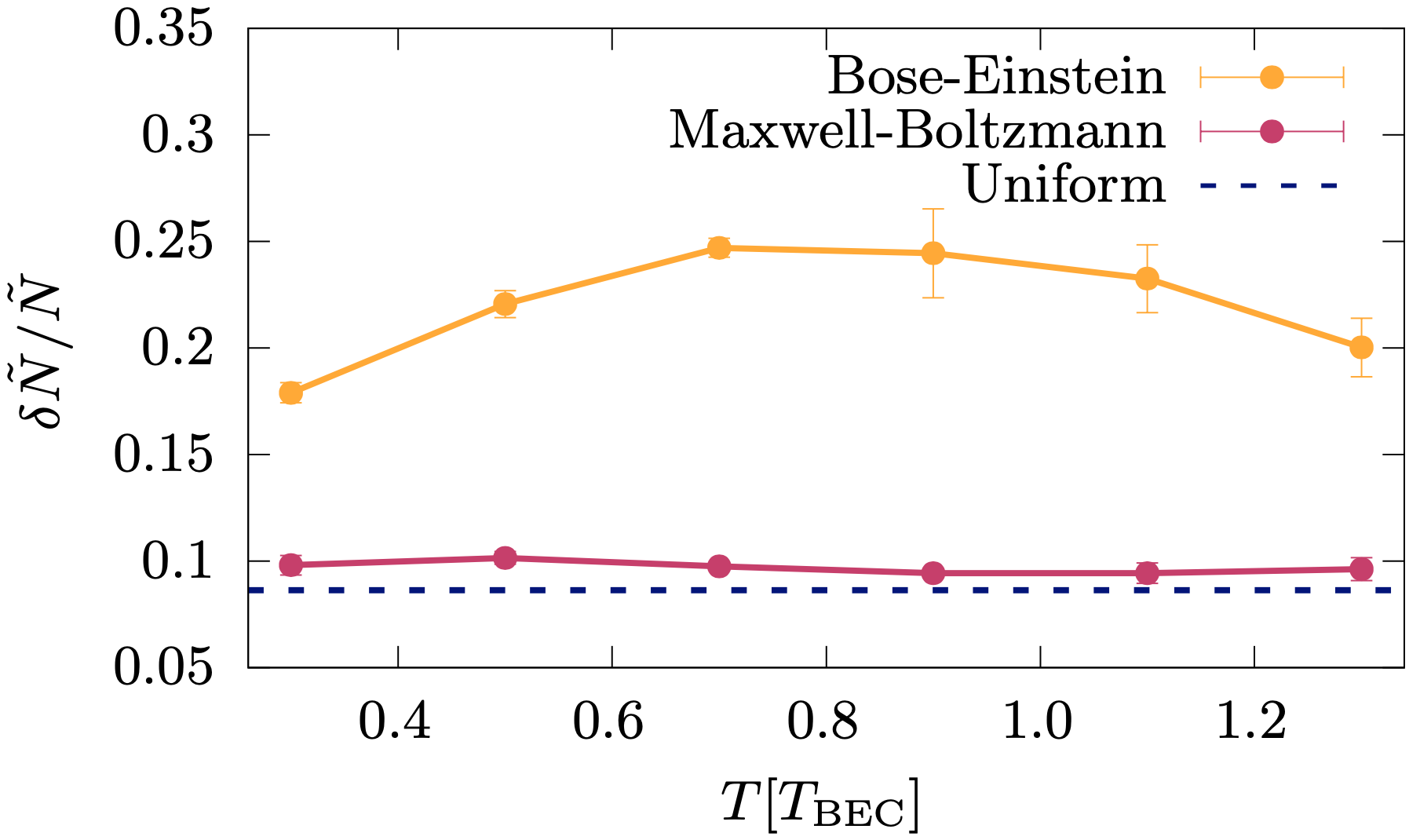}
	\caption{Local population imbalance as a function of temperature for two different statistics at $g_{12}/g = 0.93$ with $N=128$ obeying Bose-Einstein and Maxwell-Boltzmann statistics. The dashed line corresponds to a simulation with particles uniformly distributed in the box.}
	\label{fig:Statistics}
\end{figure}

\begin{figure*}[t]
	\centering
	\includegraphics[width=15cm,height=6.8cm]{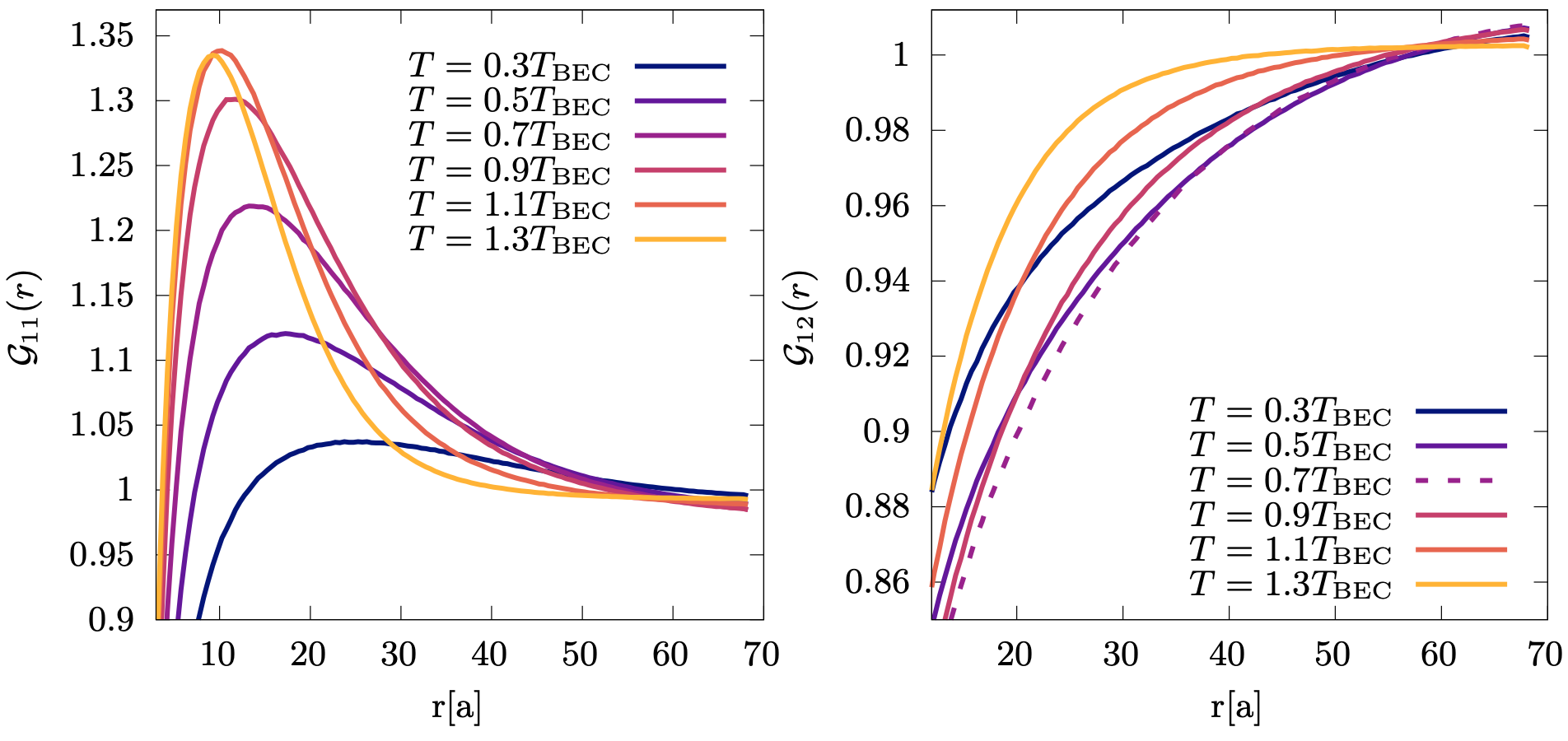}
	\caption{Pair distribution functions between particles of the same species (left) and of different species (right) at $g_{12}/g = 0.9$ with $N=256$ particles. These results have been computed in the same way as in Ref.~\cite{Spada2022} with periodic boundary conditions.}
    \label{fig:PairCorrelationFunction}
\end{figure*}

In line with Fig.~\ref{fig:DensityProfiles}, Fig.~\ref{fig:Imbalance} shows the corresponding behavior of the integrated local population imbalance $\delta\tilde{N}/\tilde{N}$ [see Eq. (\ref{deltaN})] as a function of the temperature. For $g_{12}/g = 1.50$ the system is fully separated in the limit $T \to 0$ and, by increasing the temperature, the local imbalance decreases monotonously, pointing to a tendency to mix. On the other hand, for values of $g_{12}/g \leq 1$, we see that the slope of $\delta\tilde{N}/\tilde{N}$ is positive at low temperature, showing a maximum imbalance at the characteristic temperature $T^{\star}\simeq0.7T_{\text{BEC}}$. By further increasing $T$, the local imbalance starts to decrease and the local separation between species progressively vanishes. This effect is more evident when $g_{12}/g$ approaches the threshold value. The results in Fig.~\ref{fig:Imbalance} correspond to different values of the coupling constants where what is modified is the scattering length $a_{12}$, keeping the same mass for all the particles. Another possibility is to consider different masses and modify the scattering lengths while keeping the same value of the coupling constants $g$ and $g_{12}$. In Ref.~\cite{Supplement}, we show the results for this second case that are identical to the ones of Fig.~\ref{fig:Imbalance} once the temperature has been properly rescaled. 

It is interesting to explore further the origin of the peak in the local population imbalance for the case of mixed Bose gases at $T=0$. To this end, we calculated the same local imbalance $\frac{\delta\tilde{N}}{\tilde{N}}$ but considering all the particles as distinguishable, that is, obeying the Maxwell-Boltzmann statistics. This is technically carried out by not sampling permutation cycles in the PIMC algorithm. In Fig.~\ref{fig:Statistics}, we compare the results of Bose-Einstein and Maxwell-Boltzmann statistics. As one can see, distinguishable particles do not show any peak in the local imbalance giving evidence of a quantum effect leaded by the Bose-Einstein statistics of particles. Furthermore, we also compare these results with the ones from randomly generated configurations where particles are uniformly distributed in the box (dashed line in Fig.~\ref{fig:Statistics}). We notice that, while both Maxwell-Boltzmann and uniformly distributed particles produce a finite value for $\delta\bar{N}/\bar{N}$ as a result of the procedure used to extract the local imbalance, this quantity does not show any peak as a function of temperature in sharp contrast with the case of Bose statistics. Furthermore, the result in Fig.~\ref{fig:Imbalance} corresponding to the case $g_{12}=0$ of independent components, displays a similar trend with $T$. Thus, a possible explanation of part of the effect points toward the bunching mechanism, which enhances short-range correlations between identical Bose particles. However, interspecies interactions within the miscible regime $g_{12}<g$ also play an important role by producing the maximum in the local imbalance.

In order to better understand the role of statistics and interactions, we calculate the pair distribution function between particles of the same species ($\mathcal{G}_{11}(r)$) and between particles of different species ($\mathcal{G}_{12}(r)$). These observables are properly defined in homogeneous systems where they only depend on the relative distance $r$ between particles and approach unity at large separations. For this reason we compute them in the bulk using periodic boundary conditions, for a system with the same total density $n$ and temperature $T$.

Fig.~\ref{fig:PairCorrelationFunction} shows the temperature dependence of $\mathcal{G}_{11}(r)$ and $\mathcal{G}_{12}(r)$ (left and right panel respectively) in the region of distances on the order of the mean interparticle separation. The bunching effect due to Bose statistics is active between particles of the same species and is responsible for the peak in $\mathcal{G}_{11}$. We notice that hard-core repulsions suppress short-distance bunching at any temperature, whereas the peak in $\mathcal{G}_{11}$ increases with $T$ and becomes narrower. This is a consequence of bunching correlations within the distance $\lambda_T$ reaching a maximum at $T_{\text{BEC}}$, while the thermal wavelength $\lambda_T=\sqrt{2\pi\hbar^2/mk_BT}$ shrinks with temperature. On the contrary, $\mathcal{G}_{12}$ shows a different behavior: up to $T\sim0.7T_{\text{BEC}}$ the curve is shifted to the right showing an increase in the repulsion between particles of different species while, for larger temperatures, the curve is shifted to the left. The combined effect of bunching and the interaction effect visible in $\mathcal{G}_{12}$ is probably responsible for the behavior seen in Fig.~\ref{fig:Imbalance}. It is worth mentioning that an indication of a maximum in the repulsive interspecies correlations at $T\sim0.7T_{\text{BEC}}$ is also provided by the minimum of the interspecies contact parameter at approximately the same temperature~\cite{Spada2022}.

\noindent \textit{Conclusions and Discussion.}
We study binary Bose mixtures using the PIMC method in a confined box geometry for different values of $g_{12}/g$. We define a local observable probing the structure of the mixture, in particular its local magnetization, for a finite number of particles. Different behaviors are found for this local population imbalance as temperature is increased: for independent mixtures, $g_{12}=0$, the imbalance increases steadily for all temperatures below $T_{\text{BEC}}$, viceversa, for interacting mixtures in the immiscible regime with $g_{12}>g$ we find a monotonous decrease. In the miscible regime, $0<g_{12}<g$, we find instead a non monotonous behavior, featuring a broad maximum around $T\sim0.7T_{\text{BEC}}$. At approximately the same temperature we also observe a maximum of the interspecies repulsion measured by the pair distribution function.

We interpret these findings as a result of thermal effects, statistics, and interactions. The case $g_{12}>g$ is dominated by interactions. At zero temperature, the mixture is fully phase separated and any finite temperature gives rise to a chemical potential gradient which makes the ground state unstable against particles of one species diffusing inside the region occupied by the other species~\cite{Roy2015}. The result is a tendency toward mixing which becomes stronger for increasing temperatures. The case $g_{12}=0$ follows instead from the bunching effect which involves mainly thermally induced statistical correlations. As it is well known, identical Bose particles tend to group together within a distance on the order of $\lambda_T$ and such bunching correlations are enhanced if the condensate is thermally depleted and $T_{\text{BEC}}$ is approached from below. The most interesting regime, $0<g_{12}<g$, features a more subtle interplay between interaction and statistical thermal effects. At any temperature $T$, the equilibrium state is a paramagnetic mixture exhibiting zero polarization in the thermodynamic limit~\cite{Spada2022}. However, a local partial imbalance between the two components can be the most probable state of a finite-size system. For such system, bunching tends to increase the imbalance with $T$, similarly to the $g_{12}=0$ case. Repulsive interactions between the two components make the effect of bunching more pronounced up to the characteristic temperature $T\sim0.7T_{\text{BEC}}$ where interspecies repulsion reaches a maximum. For higher temperatures the extra bunching provided by $g_{12}$ is reduced and the behavior of the local imbalance resembles again the case of independent components.

Previous works found a similar local partial separation in other quantum systems, such as a gas in a disk-shaped harmonic trap~\cite{Ma2004} or a gas with dipolar interactions~\cite{Jain2011} but neither of them reported a maximum in the local imbalance at a certain temperature below $T_{\text{BEC}}$. Furthermore, although the bunching effect is a well-known feature in Bose gas, the non monotonous behavior of the interspecies pair distribution function $\mathcal{G}_{12}(r)$ in a binary mixture is something new that can lead to interesting phenomena. For all these reasons, more research on this topic needs to be conducted and, specifically, experiments on binary Bose mixtures can shed some light on the interplay between statistics, quantum degeneracy, and interactions. The present experimental technology for producing  box potentials~\cite{Navon_naturephysics} could be an ideal setup for exploring this deep quantum phenomena. 

 \noindent \textit{Acknowledgments.}
This work has been supported by the Spanish Ministry of University under the grant FPU No. FPU20/00013, the Spanish Ministry of Economics, Industry and Competitiveness under grant No. PID2020-113565GB-C21. G.S., S.G. and S.P. acknowledge the Italian Ministry of University and Research under the PRIN2017 project CEnTraL 20172H2SC4. S.P. also acknowledges support from the PNRR MUR project PE0000023-NQSTI and the one from PRACE, for awarding access to the Fenix Infrastructure resources at Cineca, which are partially funded from the European Union Horizon 2020 research and innovation programme through the ICEI project under the grant agreement No. 800858. S.G. acknowledges also co-funding by the European Union NextGenerationEU. Views and opinions expressed are however those of the authors only and do not necessarily reflect those of the European Union or the European Research Council. Neither the European Union nor the granting authority can be held responsible for them.


\bibliography{main}

\end{document}


\preprint{APS/123-QED}

\title{Supplemental Material: Thermal-induced Separation in Repulsive Two-component Bose Mixtures}

\author{G. Pascual$^1$, G. Spada$^2$, S. Pilati$^{3,4}$, S. Giorgini$^2$ and J. Boronat$^1$}

\affiliation{
$^1$ Departament de F\'{i}sica, Universitat Polit\`ecnica de Catalunya, Campus Nord B4-B5, E-08034, Barcelona, Spain \\
$^2$ Dipartimento di Fisica, Universit\`a di Trento and CNR-INO BEC Center, 38123 Povo, Trento, Italy\\
$^3$ School of Science and Technology, Physics Division, Universit\`a di Camerino, 62032 Camerino, Italy\\
$^4$ INFN-Sezione di Perugia, 06123 Perugia, Italy
}

\date{\today}

\maketitle


\onecolumngrid

\section{Imbalanced mass ratios}
\begin{figure}[h]
	\centering
    \includegraphics[width=8.6cm,height=7.8cm]{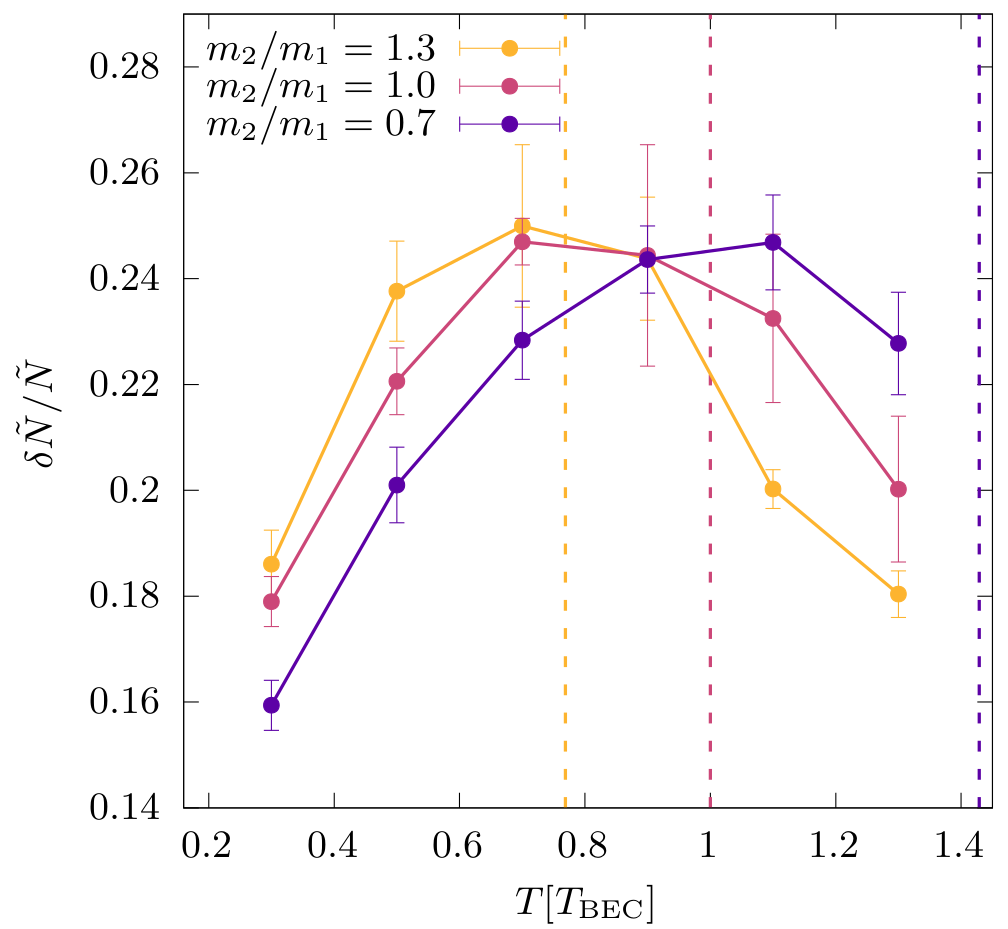}
	\caption{Local population imbalance as a function of temperature for different mass ratios. The vertical dashed lines correspond to the $T_{\text{BEC,2}}$.}
	\label{fig:ImbalancedMassRatio}
\end{figure}

When we increase the mass of the particles of the second species, we reduce its critical temperature ($T_{\text{BEC},2}= \frac{m_1}{m_2} T_{\text{BEC}}$) and, as a result, the maximum in the polarization is shifted to lower temperatures (see Fig.~\ref{fig:ImbalancedMassRatio}). This can be understood as, when one of the two critical temperatures is changed, the thermal window where the two species remain in a BEC phase and, thus, feeling the bunching effect is also modified (shortened/lengthened in the yellow/purple case in Fig.~\ref{fig:ImbalancedMassRatio}.) 

